
\input harvmac.tex
\def\at{{\tilde \alpha}}
\def\a{{\alpha}}
\def\pt{{\tilde \psi}}
\def\p{{\psi}}

\lref\dbranes{J.~Dai, R.~G.~Leigh and J.~Polchinski, Mod.~Phys.~Lett.
A4 (1989) 2073; R.~G.~Leigh, Mod.~Phys.~Lett. A4 (1989) 2767.}
\lref\afflud{
I.~ Affleck and A.~W.~W. Ludwig, Phys. Rev. Lett. {\bf 67} (1991)
161. }
\lref\bi{E.~S.~Fradkin and A.~A.~Tseytlin, Phys. Lett. 163B (1985) 123;
A.~Abouelsaood, C.~G.~Callan, C.~R.~Nappi and S.~A.~Yost, Nucl. Phys. B280
(1987) 599.}
\lref\willy{
W.~Fischler, S.~Paban and M.~Rozali, Phys.
Lett. B352 (1995) 298. }
\lref\duality{C.~M.~Hull and P.~K.~Townsend,
Nuc.~Phys. B438 (1995) 109; E. Witten, Nuc.~Phys. B443 (1995) 85;
J.~Schwarz, Lett.~Math.~Phys. 34 (1995) 309;
A.~Sen, Phys.~Lett.~B329 (1994) 217; J.~A.~Harvey and A.~Strominger,
Nuc.~Phys. B449 (1995) 535.}
\lref\integral{C. Callan, A. Felce and D. Freed, Nuc.~Phys.~B392 (1993) 551.}
\lref\cklm{C.~G.~Callan and I.~R.~Klebanov, Phys.~Rev.~Lett. 72 (1994)
1968;
C.~G.~Callan, I.~R.~Klebanov, A.~W.~W. Ludwig and J.~Maldacena,
Nuc.~Phys. B422 (1994) 417;
C.~G.~Callan, I.~R.~Klebanov, J. Maldacena and A. Yegulalp,
Nuc.~Phys. B443 (1995) 444.}
\lref\polone{J.~Polchinski, ``Dirichlet-Branes and Ramond-Ramond Charges'',
NSF Institute for Theoretical Physics preprint, hep-th/9510017.}
\lref\poltwo{ J.~Polchinski, Phys.~Rev.~D50 (1994) 6041.}
\lref\clnyone{C.~Callan, C.~Nappi, C.~Lovelace and S.~Yost, Nuc.~Phys.~B308
(1988) 221.}
\lref\cmp{C.~Callan, J.~Maldacena and A.~Peet, ...}
\lref\miao{M.~Li, ``Boundary States of D-Branes and Dy-Strings'',
Brown University preprint, hep-th/9510161.}
\lref\ewone{E.~Witten, ``Bound States of Strings and p-Branes'',
IAS preprint, hep-th/9510135.}
\lref\iglar{I.~R.~Klebanov and L.~Thorlacius, ``The Size of p-Branes'',
Princeton preprint PUPT-1574, hep-th/9510200.}
\lref\jsdual{J.~Schwarz, ``An SL(2,Z) Multiplet of Type IIB Superstrings''.
California Institute of Technology preprint, hep-th/9508143.}
\lref\bachas{C.~Bachas, ``D-Brane Dynamics'', ITP and Ecole Polytechnique
preprint, hep-th/9511043.}
\lref\cmp{C.~Callan, J.~Maldacena and A.~Peet, ``Extremal Black Holes
as Fundamental Strings'', Princeton preprint PUPT-1565, hep-th/9510134.}

\Title{\vbox{\baselineskip12pt
\hbox{PUPT-1578}\hbox{hep-th/9511173}}}
{\vbox{\centerline{D-Brane Boundary State Dynamics}}}
\centerline{Curtis G. Callan, Jr., and Igor R. Klebanov}
\centerline{\it Department of Physics, Princeton University}
\centerline{\it Princeton, NJ 08544}
\vskip.3in
\centerline{\bf Abstract}
We construct the open string boundary
states corresponding to various time-dependent deformations of the
D-brane and explore several ways in which they may be used to study
stringy soliton collective coordinate quantum dynamics. Among other
things, we find that D-strings have exact moduli corresponding to
arbitrary chiral excitations of the basic soliton. These are presumably
the duals of the BPS-saturated excitations of the fundamental Type IIB string.
These first steps in a systematic study of the dynamics and interactions
of Dirichlet-brane solitons give further evidence of the consistency of
Polchinski's new approach to string soliton physics.
\Date{11/95}
\eject

\newsec{Introduction}

Polchinski's remarkable proposal \polone\ that the R-R charge-carrying
solitonic states of Type II string theory can be given an exact conformal
field theory description via open strings with Dirichlet boundary conditions
\dbranes\ has opened up a new chapter in the development of string theory. The
problem of the proper description of solitons in string theory has always
been fascinating, if elusive, but it has recently become particularly
urgent with the realization that apparently different pairs of string theories
are dual to each other, with the solitons of one being the fundamental
strings of the other \duality. Polchinski suggests that, in order to
obtain a full string theory description of many solitons, rather than look
for some complicated conformal field theory of an extended spacetime object,
it suffices to consider an
open string subject to just enough Dirichlet boundary
conditions to localize the center of mass of the soliton. This proposal,
shocking in its simplicity (at least to anyone who has struggled with the
more conventional approach to string solitons), has already passed quite
a few consistency tests bearing mainly on static properties of the
solitons \refs{\miao,\ewone,\iglar}.
In this paper, we examine how the new proposal deals with some
issues in soliton dynamics, {\it viz.} collective coordinate quantization,
scattering and excitation of internal degrees of freedom.

Once the soliton problem is recast as a problem in open string dynamics, the
technical issue becomes that of finding an appropriate conformally invariant
``boundary state'' \clnyone\ and extracting target space information, such as
the soliton mass, from it. As an aside, we note that one of the more puzzling
old-style string soliton problems was to find a conformal field theory
construction of the soliton mass: It is not the central charge and the
correct construction, discovered only recently, is quite subtle \willy.
Open string boundary conformal field theories, however, do
define a quantity analogous to the central charge, known as the zero
temperature boundary entropy \afflud, which has all the right scaling
properties to be a target space action density. Polchinski's
merging of solitons with open strings empowers us to make such an
identification \poltwo\ and one of our concerns
will be to show how it works in dynamical detail.

More generally, in this paper we will construct the boundary states which
describe solitons in various states of collective coordinate excitation
and will use these objects to extract as much dynamical information as we can.

\newsec{Moving the D-Brane Boundary State}

In order to endow a general Dirichlet D-brane with interesting
dynamics one introduces the following boundary action
\refs{\dbranes,\ewone,\miao}:
$$ S_b = \oint d\sigma \left [ \sum_{i=0}^p A_i (X^0, \ldots, X^p)
{\partial\over \partial \sigma} X^i
+ \sum_{j=p+1}^{9} \phi_j (X^0, \ldots, X^p)
{\partial\over \partial \tau} X^j \right ]~.
$$
We have suppressed the fermionic terms associated with world sheet
supersymmetry and we have taken the boundary to lie at constant $\tau$.
The $A_i$ are gauge fields on the D-brane world volume which need to be
turned on if we wish to give it some NS-NS charge. The fields
$\phi_j$, on the other hand, are scalars from the
world volume point of view: they describe transverse motions of the
D-brane. The boundary gauge fields are of course not arbitrary:
for string theory consistency, they must define a boundary conformal
field theory. In this paper we will consider some very simple, physically
interesting, special choices which are manifestly conformal.
While the above action has no symmetry between $X^i$, $i=0, \ldots, p$
and $X^j$, $j=p+1, \ldots, 9$, the symmetry is restored if we perform
a T-duality transformation on $X^j$. This is why the D-brane dynamics
has some hidden simplicity.

For definiteness we will construct boundary states for
charged moving $1$-branes, although our techniques may be
generalized to any D-brane. For the $1$-brane,
the coordinates $X^0$ and $X^1$, as well as their fermionic partners,
have  Neumann boundary conditions, while
$X^j$ ($j=2, \ldots, 9$) and their fermionic partners
have Dirichlet boundary conditions. The NS-NS part of the
stationary uncharged membrane's boundary state is,
$$ | B_0\rangle =
\prod_{j=2}^{9} \delta(X^j) | B_\alpha\rangle | B_\psi\rangle
| B_{\rm gh} \rangle
$$
The explicit expressions for the factors can be obtained from the
results of \clnyone\ by applying the T-duality transformation needed
to impose the appropriate Dirichlet boundary conditions:
$$\eqalign{& | B_\alpha\rangle=
\exp~\bigl\{ \sum_{n=1}^\infty {1\over n} ( \at^0_{-n} \a^0_{-n}
- \at^1_{-n} \a^1_{-n} + \at^j_{-n} \a^j_{-n} )
\bigr\} |0\rangle \ , \cr
& | B_\psi\rangle =
\exp~\bigl\{ \sum_{n>0} ( \pt^0_{-n} \p^0_{-n}
- \pt^1_{-n} \p^1_{-n} + \pt^j_{-n} \p^j_{-n} )
\bigr\} |0\rangle \ , \cr
& | B_{\rm gh} \rangle =
\exp~\bigl\{ \sum_{n>0} ( \tilde\gamma_{-n} \beta_{-n}-
\gamma_{-n} \tilde \beta_{-n}) + \sum_{m=1}^\infty
(c_{-m} \tilde b_{-m} + \tilde c_{-m} b_{-m} )
\bigr\} |Z\rangle \ , \cr
} $$
where $|Z\rangle$ is the appropriate ghost vacuum.
The mode numbers of the fermions and the superghosts are half-odd-integral.

Now introduce an electric field in the $1$-direction,
and also give the D-brane some transverse velocity in the
$2$-direction. To this end we introduce the boundary interaction
$$ \oint d\sigma \left [
E X^0 {\partial\over\partial\sigma} X^1
+ V  X^0 {\partial\over\partial t} X^2 \right ] ~,
$$
together with appropriate fermionic terms required by the world sheet
supersymmetry.
The primary effect of this quadratic interaction is to produce a
Lorentz boost on the left-moving part of the boundary state relative
to the right-moving part. A secondary, but crucial,
effect of turning on the boundary gauge fields $A_i$ and $\phi_j$ is the
appearance of an overall normalizing factor in front of the boundary state.
As shown in \refs{\bi,\clnyone},
this factor is the Born-Infeld action for the boundary
gauge fields. In the case at hand, where the boundary field strengths are
$F_{01}=E$, $F_{02}=V$, the Born-Infeld normalizing factor reduces to
$\sqrt{1-E^2-V^2}$ (the Lorentz signature of the
metric is of course crucial here). The Born-Infeld action term is not
affected by the Dirichlet boundary conditions for some of the
fields \dbranes,
but the operator structure of the boundary state does suffer a rather
trivial modification which we will make explicit and comment on shortly.

The explicit form of the boosted boundary state generated by turning on
both $E$ and $V$ is
$$
|B_{E, V}\rangle = \sqrt{1-E^2-V^2}\ \exp(O) |B_0\rangle
$$
where $O= O_\p + O_\a$ with
$$
O_\a= 2\delta\left [ {E\over \sqrt{E^2+V^2}}
\sum_{n>0} {1\over n} ( \a^0_{-n}\a^1_n
-\a^0_n\a^1_{-n})+ {V\over \sqrt{E^2+V^2}}
\sum_{n>0} {1\over n} ( \a^0_{-n}\a^2_n
-\a^0_n\a^2_{-n})\right ]
$$
and $O_\p$ generates the same boost on the fermions.
The hyperbolic angle $\delta$ is related to the ``velocity'' $(E, V)$
in the usual manner of Lorentz tranformations:
$\cosh\delta = {1/\sqrt {1-E^2-V^2}}$. The net
effect is a boost of the left-movers relative to the right-movers.
This expression for the boundary state reveals the interesting fact that,
while the boundary state is not $SO(1, 9)$ invariant, the operator
that generates it from the ``bare'' boundary state is an element of $SO(1, 9)$.

Consider first the simple special case $V=0$ (so that
$\cosh\delta = {1/\sqrt {1-E^2}}$).
The dynamic boundary state is constructed from the static one by replacing
$$
\pmatrix { \a^0\cr \a^1\cr } \rightarrow
\pmatrix { \cosh (2\delta) & \sinh (2\delta)\cr
\sinh (2\delta) & \cosh (2\delta)\cr } \pmatrix { \a^0\cr \a^1\cr }
\ , \qquad
\pmatrix { \p^0\cr \p^1\cr } \rightarrow
\pmatrix { \cosh (2\delta) & \sinh (2\delta)\cr
\sinh (2\delta) & \cosh (2\delta)\cr } \pmatrix { \p^0\cr \p^1\cr }
$$
and multiplying by the Born-Infeld factor $\sqrt{1-E^2}$. In terms of $E$,
the boost matrix is
$$ \pmatrix { \a^0\cr \a^1\cr } \rightarrow
{1\over 1-E^2} \pmatrix { 1+E^2 & 2E\cr 2E & 1+E^2\cr}
\pmatrix { \a^0\cr \a^1\cr }
\ .$$
By expanding the boundary state in creation operators, one can read off
that the source for the antisymmetric tensor $B_{01}$ is $2E/\sqrt {1-E^2}$,
while the graviton source strength is $(1+E^2)/\sqrt {1-E^2}$. In the
following sections, we will see that various conclusions about soliton
dynamics can be read off from these results.

\newsec{Quantizing the Born-Infeld Action}

In order to relate $E$ to the NS-NS charge and to identify the energy
per unit length of the string,  we need an action. We will make the plausible
guess that the normalization factor for the boundary state (the same
thing as the disk amplitude) can be taken as the action for the collective
coordinate $A_1 (t)$ and quantized. To give the string a definite length,
let us compactify $X^1$ on a circle of length $l$.
Now the Born-Infeld
Lagrangian collapses to
$$
L=-{l\over \lambda}\sqrt {1- \dot A_1^2}\ , \qquad \dot A_1 = E~.
$$
For completeness, we have introduced the power of the string coupling constant
$\lambda$ appropriate to the origin of this action in the disk amplitude
(had the action somehow been derived from a sphere amplitude, the correct power
would have been $\lambda^{-2}$).

A subtle but crucial feature of the theory is that
the presence of large gauge transformations makes $A_1$ a compact variable
with period $2\pi/l$. Therefore,
the momentum conjugate to $A_1$,
$$
Q= {l\dot A_1\over \lambda\sqrt {1- \dot A_1^2} }
\ ,$$
is quantized in units of $l$.\foot{Our definition of the charge differs from
that in \miao. }
The Hamiltonian derived from $L$ is
$$
H= Q \dot A_1 -L =\sqrt {\left ({l\over \lambda}\right )^2+ Q^2}=
l\sqrt {\left ({1\over \lambda}\right )^2+ n^2}\ ,
$$
where $n=Q/l$ is an integer. Remarkably, the string energy per unit length
is precisely the BPS formula for the tension of $(n, 1)$ strings
\refs{\jsdual,\ewone,\miao}!
The boundary state can now be rewritten in terms of the charge, rather than
the less directly meaningful field strength $E$ and we find, for example,
that the source for $B_{01}$ is $2n$, which is proportional
to the NS-NS charge per unit length.

We can give a similar treatment to the general case of a charged, moving
$1$-brane. Here both $E$ and $V$ are non-vanishing and the boundary state
is defined by a more complicated Lorentz transformation,
one which mixes the
longitudinal and transverse directions:
\eqn\biglt{
\pmatrix { \a^0\cr \a^1\cr \a^2\cr} \rightarrow
{1\over 1-E^2-V^2} \pmatrix { 1+E^2+V^2 & 2E &2V\cr 2E & 1-V^2+E^2&
2VE\cr 2V & 2VE & 1-E^2+V^2}
\pmatrix { \a^0\cr \a^1\cr \a^2\cr}
}
and similarly for the left-moving fermion modes.
The Born-Infeld action normalizing factor, which we want to treat as
the effective lagrangian for the collective coordinates, now turns out to be
$$
L=-{l\over\lambda}\sqrt {1- \dot A_1^2- \dot X_\perp^2}
$$
and the Hamiltonian is
$$
H=\sqrt {\left ({l\over \lambda}\right )^2+ Q^2 + P_\perp^2}
$$
where
$$ Q= {l\dot A_1\over \lambda \sqrt {1- \dot A_1^2- \dot X_\perp^2}}\ ,
\qquad P= {l\dot X_\perp\over \lambda\sqrt {1- \dot A_1^2- \dot X_\perp^2}}~.
$$
The string energy per unit length is
$$ \sqrt {\left ({1\over \lambda}\right )^2+ n^2 + p_\perp^2}
$$
where $p_\perp= P_\perp/l$ is the momentum density.
This is just the relativistic expression for the energy density of a moving
straight string! If we examine
the boundary state appropriate to this case, we see that the
source for $B_{01}$ is again equal to $2n$. There is also a source for
$B_{12}$ equal to $-2Vn$, but no source for $B_{02}$.

Next, we might ask whether there is an action to quantize in order to
obtain the tension of the multiply-wound $(n, m)$ strings, where
$m$ refers to the R-R charge and $n$ to the NS-NS charge.
For the doubly-wound strings
($m=2$), for instance, the BPS mass formula tells us that such states
really exist only for $n$ odd (the even-charged states are neutrally
stable with respect to splitting into two singly-wound strings).
Is there a generalization
of the Born-Infeld action which allows us to examine this case too?
Witten has shown \ewone\ that the proper setup for discussing the dynamics
of multiple strings is a $N=8$ supersymmetric $U(m)$ gauge theory in
$1+1$ dimensions. He has further argued that there are certain vacua
where the $SU(m)$ part of the theory develops a mass
gap (they are described by placing $n$ quarks at infinity
such that $m$ and $n$ are relatively prime).
A physical consequence of this mass gap is the formation of
a bound state. If the $SU(m)$ part of the gauge field is frozen out,
then we may replace the $U(m)$ gauge field $\bf A_\mu$ by $A_\mu I$, where
$I$ is the $m$-dimensional identity matrix.
Now the relevant
Born-Infeld lagrangian is
$$
L=-{lm\over \lambda}\sqrt {1- \dot A_1^2}\ ,
$$
where the factor of $m$ comes from tracing over the
$U(m)$ indices.
Quantization of this $U(1)$ theory is analogous to what we encountered
in the $m=1$ case. All the formulae carry over with the replacement
of $1/\lambda$ by $m/\lambda$.
Thus, we find that the tension of $(n, m)$ strings is
$$\sqrt {\left ({m\over \lambda}\right )^2+ n^2}\ ,
$$
Remarkably, this again agrees with the BPS formula!
While this is very encouraging, we clearly need a better understanding of
why the $SU(m)$ part of the theory may be ignored when
$m$ and $n$ are relatively prime.

\newsec{Forces On Moving D-Branes}

Boundary states can be used to calculate the forces between D-branes.
Roughly speaking, the annulus amplitude obtained by gluing two boundary
states together gives the interaction energy due to closed string exchange.
The first thing we would like to extract from such an exercise is that the
force between separated, but otherwise identical, D-branes vanishes. This
is the no-force condition on BPS saturated states and should follow from
supersymmetric boson-fermion cancellations in the annulus amplitude. The
requisite cancellations have been shown to occur for the basic soliton \polone,
and for static charged solitons \miao. We will verify that the force
cancellation also occurs for charged, translating solitons. The same formalism
allows us to calculate the force between solitons of different charges and/or
non-zero relative velocity. In these cases the BPS saturation argument for
vanishing force doesn't work and we indeed find a non-vanishing long-range
force with an interesting dependence on charges and velocities.

This calculation brings the full boundary state, including both the NS-NS
and R-R
sectors, into play. This object was constructed in studies of the
Fischler-Susskind mechanism in string theory \clnyone\ and Li has recently
shown how to modify these old results to obtain the D-brane boundary
state \miao. We will make heavy use of his results here.

The boundary states are a convenient tool for summing over the forces
mediated by the NS-NS and the R-R bosons. For simplicity, we will restrict
our attention to the long range force and project the boundary state
onto the massless levels. In the picture with superghost charge $-2$,
the metric and the antisymmetric tensor components of that projection
may be concisely written as
$$ \sqrt{1-E^2 - V^2}\ \pt^\mu_{-1/2}
       (\eta\cdot\Lambda_{E, V}\cdot T)_{\mu\nu}\p^\nu_{-1/2} |Z\rangle
$$
where $\Lambda_{E, V}$ is the ten-dimensional Lorentz transformation
matrix whose non-trivial $3\times 3$ corner is given in \biglt,
$\eta = {\rm diag} (-1, 1, \ldots, 1)$ is the Minkowski metric and
$T = {\rm diag} (1, 1, -1, \ldots, -1)$ is the net effect of the T-duality
transformation which imposes the Dirichlet boundary condition on
$X^j$, $j=2,\ldots, 9$. A ghost dilaton component
$$
\sqrt{1-E^2 - V^2}\ (\beta_{-1/2}\tilde\gamma_{-1/2} -
\tilde\beta_{-1/2}\gamma_{-1/2} ) |Z\rangle
$$
must be added to this to obtain the full boundary state.

We will use these results to calculate the long-range force between two
$1$-branes carrying different charges and moving with different velocities.
By inserting the closed string propagator between an $(E_1, V_1)$
boundary state and an $(E_2, V_2)$ boundary state, we find that the
net interaction due to the NS-NS exchanges is equal to that between two $E=0$
stationary $1$-branes multiplied by the prefactor
$$
{\cal P_{NS}} = \sqrt{1-E_1^2 - V_1^2} \sqrt{1-E_2^2 - V_2^2}\
\bigl [ \tr (\eta~\Lambda^{t}_{E_2, V_2} ~\eta~\Lambda_{E_1, V_1}) -2]~.
$$
The basic interaction potential due to massless exchanges in eight
transverse dimensions is $r^{-6}$, and we could, if needed, obtain its
absolute normalization.

Space-time supersymmetry leads us to expect that, for two stationary $E=0$
D-branes, the net NS-NS force is exactly cancelled by the net R-R
force. The general R-R boundary state is quite awkward to construct, but
there is a simple recipe for the projection of the constant gauge field
strength boundary state onto the massless levels. We will just state the recipe
and refer the interested reader to \refs{\clnyone,\miao}\ for detailed
justification. The first observation is that this projected boundary
state involves only world sheet fermion zero modes (there is not even a
normalizing determinant factor: the bosonic boundary determinant exactly
cancels its fermionic partner). The zero-mode part of the constant gauge
field strength boundary state is represented by a polynomial in components
of $F_{\mu\nu}$ with the spacetime indices saturated on zero modes
$\theta_0^\mu=\psi_0^\mu +\tilde\psi_0^\mu$.
Because of anticommutativity, no individual $\theta_0^\mu$
can appear more than once in any monomial. For the case at hand we have
$$
   {\cal P}_{R}\ =\ 1 - E \theta_0^0\theta_0^1 - V \theta_0^0\theta_0^2 ~.
$$
The three terms $1$, $\theta_0^0\theta_0^1$ and
$\theta_0^0\theta_0^2$ can be regarded
as orthogonal states of unit normalization in computing inner products
between two boundary states. It follows that the net force due to the R-R
exchanges is equal to the corresponding
force between two $E=V=0$ $1$-branes multiplied
by $1- E_1 E_2 - V_1 V_2 $ (the minus signs come from the anticommutativity
properties of the $\theta_0^\mu$). The overall normalization of the force
will contain an important factor $8$ which comes from the dimensionality
of the spinor space on which the fermion zero modes act.

Adding up the NS-NS and the R-R forces we get the following
prefactor for the total long-range force,
\eqn\force{ {\cal P}_{tot} =
\sqrt{1-E_1^2 - V_1^2} \sqrt{1-E_2^2 - V_2^2}\
\bigl [ \tr (\eta~\Lambda^{t}_{E_2, V_2} ~\eta~\Lambda_{E_1, V_1}) -2]
                      - 8 (1- E_1 E_2 - V_1 V_2)  ~.}
Using the defining relation of Lorentz tranformations, it is
a simple matter to check that the total force between identical D-branes
($E_1= E_2$ and $V_1=V_2$) cancels. This is a consequence of BPS saturation
which, in turn, is a consequence of the fact that the boundary state, even
for non-zero $E$ and $V$, is annihilated by half the supersymmetries.
Evaluating \force\ for general $E$ and $V$ gives
\eqn\total{ \eqalign{
{\cal P}_{tot} = &
4\ {2 - (E_1+E_2)^2 - (V_1+V_2)^2 +(E_1^2+ V_1^2) (E_2^2+ V_2^2)
  + (E_1 E_2+ V_1 V_2)^2 \over \sqrt{1-E_1^2 - V_1^2}
\sqrt{1-E_2^2 - V_2^2}}\cr
           & \qquad - 8 (1- E_1 E_2 - V_1 V_2)  }
}
For $E_1=E_2$, this expression describes the velocity-dependent force
between two identical D-branes (for which the static force vanishes by
BPS saturation). The term $\sim (V_1-V_2)^2$, in turn, is the metric on
moduli space for the scattering of two D-branes. It is
a simple calculation to show that, for $E_1=E_2$ this vanishes,
and the leading term in the force is $\sim (V_1-V_2)^4$
(this agrees with Bachas's conclusion \bachas\
that the metric on the moduli space of two $1$-branes vanishes).
Remarkably, the forces between long fundamental strings have the
same type of velocity dependence for small velocities \cmp!
This might serve as another argument in favor of the $SL(2, Z)$ duality
relating the solitonic and fundamental strings \jsdual.
Furthermore, \total\ shows that for $1$-branes with unequal
NS-NS charges the metric on moduli space no longer vanishes
(the same property holds for the long fundamental strings).

\newsec{Waves on D-branes}

Now that we have found a boundary CFT description of D-branes
rigidly moving in transverse directions, it is interesting to look for
their internal excitations. In the following we will work out
a very interesting example of a $1$-brane stretched around a compact
dimension of length $l$. Generalizations to D-branes with $D >1$ are
possible and will be discussed briefly.

In order to describe a Dirichlet string stretched around a compact
dimension, we identify $X^1\sim X^1 + l$. A transverse excitation,
polarized in the direction $\epsilon^j$ and left-moving along the string,
is described by the boundary operator
\eqn\wave{ \oint d\sigma  \left [
\sum_p a_p e^{ip X^+}\epsilon^j (
{\partial\over \partial \tau} X^j + ip \p^+ \p^j) + c. c.
\right ]}
where we have introduced the light-cone components,
$$
X^\pm = X^0 \pm X^1\ , \qquad \p^\pm = \p^0 \pm \p^1
$$
Note that the compactness of $X^1$ restricts the allowed values of
$p$ to $2\pi n/l$, where $n$ is an integer. The crucial feature of the
operator above is that, due to the Minkowski signature of space-time,
it has the marginal dimension 1 for any value of $p$.

We will now argue that a theory with the boundary operator \wave\
added to the action describes an exactly conformal theory for
any $a_p$. The first step towards establishing this is the
calculation of the disk partition function $Z_{\rm disk}$. For instance,
the term of order $(a_p a^\star_p)^{\ m}$ in the expansion of
$\ln Z_{\rm disk}$ is given by the connected correlation function of $2m$
boundary operators. Remarkably,
all the  connected $n$-point functions of the operator \wave\ vanish
for $n>2$.
This is because the two-point functions of the light-cone components
obey
$$ \langle X^+ (\sigma_1) X^+ (\sigma_2)\rangle =
\langle \p^+ (\sigma_1) \p^+ (\sigma_2)\rangle = 0
\ .$$

A simple way to confirm that the operator \wave\
is exactly marginal is by adopting the light-cone
gauge $X^+=\tau$. In this context
the operator \wave\ implements a shift of the transverse coordinates
which depends only on $\tau$. Since the light-cone boundary state
is constructed at fixed $\tau$, the only effect
on the light-cone boundary state is a shift in the zero mode,
which implies that the theory remains free.
The conclusion is that an $arbitrary$
left-moving excitation of the Dirichlet $1$-brane is a conformally
invariant background.

The next step would be to show that this class of boundary conformal
field theories can be extended to a full supersymmetric boundary state,
annihilated by a linear combination of the left- and right-moving
supersymmetry generators. That would show that these propagating wave
deformations of the $1$-brane are still BPS-saturated states. We know how
to do this supersymmetric extension for the constant background gauge
field case and we believe that it can be done here as well (we'll save the
details for another paper). These states are presumably the duals
of the known BPS-saturated excited states of an extended Type IIB fundamental
string. The latter are constructed by applying purely left-moving, or
purely right-moving oscillators to the fundamental string ground state
and correspond to disturbances propagating in only one direction along the
string. The left-right level matching condition then requires that the
winding fundamental string be endowed with a non-vanishing longitudinal
momentum in the winding direction ($p_R^2 - p_L^2= N_L-N_R$ to be precise).
There should be a corresponding condition on the $1$-brane excitations,
but we have yet to identify it.

Now we turn to
boundary actions which describe a Dirichlet string carrying
both left- and right-moving excitations:
\eqn\newwave{ \oint d\sigma \left [
\sum_p a_p e^{ip X^+}\epsilon^j (
{\partial\over \partial \tau} X^j + ip \p^+ \p^j)
+\sum_k \tilde a_k e^{ik X^-} \tilde \epsilon^j (
{\partial\over \partial \tau} X^j + ik \p^- \p^j ) + c. c.
\right ]}
It is easy to see that, unlike \wave,
in general this does not define a conformal field theory.
Consider, for instance, the connected four-point function for the
forward scattering of two right-movers and two left-movers.
This is the coefficient of the $a_p a^\star_p a_k a^\star_k$ term in the disk
amplitude. If this four-point function does not vanish,
then the theory has a non-trivial beta function.
Let us choose $\epsilon^j \tilde\epsilon_j=0$,
so that the right and left-movers are polarized at right angles.
The four-point function is then given by the following integral,
$$ \int_{-\infty}^\infty dt
\left (|t|^{4pk} |1-t|^{-4pk}- 1\right )-
\int_{-\infty}^\infty dt~ |t|^{4pk} |1-t|^{-4pk}
{4pk\over t(t-1)}
$$
This integral can be calculated exactly and exhibits
an infinite sequence of poles in the variable $4pk$.
The origin of these poles, which are located at
the odd integer values of $4pk$,
may be traced to massive string states
propagating along the $1$-brane.
Consider, for example, the OPE
$$ \eqalign{
e^{ip X^+}{\partial\over \partial \tau} X^2 (\sigma_1)\ &
e^{ik X^-}{\partial\over \partial \tau} X^3 (\sigma_2) \cr
&\sim |\sigma_1-\sigma_2|^{-4pk} {\partial\over \partial \tau} X^2
{\partial\over \partial \tau} X^3 e^{i (p+k) X^0 + i (p-k) X^1} +\ldots}
$$
This identifies the operator which gives rise to the first pole
in the collision of a right-mover and a left-mover, located at $4pk=1$.

Thus, we reach a conclusion that the Dirichlet strings differ from the
fundamental strings in an essential way: while
the latter are exactly described by the
Nambu-Goto action, for the former it is at best an approximate low-energy
effective description.
The presence
of the infinite sequence of poles indicates that there are
extra degrees of freedom on the Dirichlet brane world volume which
are not contained in the Nambu-Goto action. These new degrees of freedom
are simply the massive modes of the open string whose ends are attached
to the $p$-brane. The effects of these modes have already been observed
in the scattering of massless particles off the $p$-branes \iglar:
they give rise to an infinite sequence of poles in the $s$-channel.
The presence of such new soliton degrees of freedom raises the
question whether the $SL(2, Z)$ symmetry,
which interchanges fundamental strings
and Dirichlet $1$-branes is really an exact symmetry of the type IIB theory.
If the $SL(2, Z)$ symmetry is exact, it requires that the fundamental
string behavior changes very much as the coupling increases.
Since the $SL(2, Z)$ relates fundamental strings at strong coupling
to the Dirichlet strings at weak coupling, our results imply that
that the strongly coupled fundamental strings
are not exactly described by the Nambu-Goto action, but
acquire an infinite set of additional degrees of freedom.
This result is not unexpected:
because of loop corrections to string equations of motion the conformal
field theory is not applicable to strongly coupled fundamental strings.

While the simultaneous presence of the left and right movers on
the $1$-brane in general destroys the world sheet conformal invariance,
left and right-movers can coexist if the
left and right-moving momenta are chosen in such a way as
to eliminate logarithmic divergences in the disk partition function. Consider,
for instance, the boundary action
\eqn\swave{ \oint d\sigma ~A~
\bigl [ e^{ip X^+}\epsilon^j (
{\partial\over \partial \tau} X^j + ip \p^+ \p^j)
+ e^{ip X^-} \epsilon^j (
{\partial\over \partial \tau} X^j + ip \p^- \p^j ) + c. c. \bigr ]
}
which describes a standing wave of amplitude $A$ on the
Dirichlet string. If $2p^2=1, 2, 3,
\ldots, $ then the perturbative expansion of the disk partition function
reduces to integrals of integer powers of $\sigma_i-\sigma_j$.
As shown in \integral, such integrals may be reduced to complex contour
integrals enclosing at worst multiple pole singularities. As a result,
they contain no logarithmic divergences and represent conformal fixed points.
Thus, it appears that we have found a whole new family of boundary
conformal field theories similar to the exact solutions
of dissipative quantum mechanics obtained in \cklm.
The new boundary CFT's are especially interesting
because they describe vibrational excitations of the Dirichlet
$1$-branes with arbitrarily large amplitude.
An even more general class of boundary CFT's would
describe the excitations of other D-branes, and we hope to
give a detailed treatment of these theories in future publications.

\newsec{Conclusions}

In this paper we have shown how Polchinski's boundary state approach
to dual string solitons can be extended to deal with the dynamical issues
of zero mode quantization, soliton scattering and excitation of internal
degrees of freedom. Everything we have found is consistent with generic
physical expectations for the properties of solitonic extended objects.
Precise predictions for the spectrum of soliton excitations can be obtained
from BPS saturation and duality with extended fundamental strings. In the
simplest case, namely the NS-NS charged excitations of the once-wound soliton
string, we could compare these predictions with the the results of quantizing
a zero-mode coordinate and found perfect agreement. Our overall conclusion
is that Polchinski's Dirichlet brane proposal contains within it the seeds
of a complete and consistent dynamics of dual solitons.

What has been done is just the tip of a large iceberg.
The boundary states corresponding to most non-trivial excited states of
the soliton have yet to be constructed in any detail. A lot
remains to be done in that direction. Also, the conformal boundary state
corresponds, roughly speaking, to the classical action and classical
field configurations of the soliton. It is absolutely crucial to quantize
this system, especially to explore duality issues, and we have shown how,
making some plausible assumptions, that can be done in the simplest case.
What's at issue more generally is the string theory analog of collective
coordinate quantization. This is already a complicated subject in field
theory and virtually nothing (with the exception of the interesting work
reported in \willy) is known about how it works in string theory.
Now that Polchinski has provided us with a manageable classical starting point
for the discussion, perhaps progress can be made.

\newsec{Acknowledgements}

We are grateful to J. Maldacena, V. Periwal, C. Schmidhuber,
L. Thorlacius and A. Tseytlin for
interesting discussions.
This work was supported in part by DOE grant DE-FG02-91ER40671.
The work of IRK was also supported in part by
the NSF Presidential Young Investigator Award PHY-9157482 and the
James S. McDonnell Foundation grant No. 91-48.

\listrefs
\bye